\documentclass{vgtc}                          

\ifpdf
  \pdfoutput=1\relax                   
  \usepackage{graphicx}                
  \DeclareGraphicsExtensions{.pdf,.png,.jpg,.jpeg} 
\else
  \usepackage{graphicx}                
  \DeclareGraphicsExtensions{.eps}     
\fi%

\graphicspath{{figures/}{pictures/}{images/}{./}} 

\usepackage{microtype}                 
\PassOptionsToPackage{warn}{textcomp}  
\usepackage{textcomp}                  
\usepackage{mathptmx}                  
\usepackage{times}                     
\usepackage{cite}                      
\usepackage{tabu}                      
\usepackage{booktabs}                  
\usepackage[normalem]{ulem} 
\usepackage{xcolor}

\onlineid{1092}

\vgtccategory{Research}

\vgtcinsertpkg

\title{Visual Analytics of Student Learning Behaviors on K-12 Mathematics E-learning Platforms }

\author{Meng Xia\thanks{e-mail: iris.xia@connect.ust.hk}\\ %
        \scriptsize Hong Kong University of \\\scriptsize Science and Technology %
\and Huan Wei\thanks{e-mail: hweiad@connect.ust.hk}\\ %
     \scriptsize Hong Kong University of \\\scriptsize Science and Technology %
\and Min Xu\thanks{e-mail: mxuar@connect.ust.hk}\\ %
     \scriptsize Hong Kong University of \\\scriptsize Science and Technology
\and Leo Yu Ho Lo\thanks{e-mail: leoyuho.lo@connect.ust.hk}\\ %
     \scriptsize Hong Kong University of \\\scriptsize Science and Technology %
\and Yong Wang\thanks{e-mail: ywangct@connect.ust.hk}\\ %
     \scriptsize Hong Kong University of \\\scriptsize Science and Technology
\and Rong Zhang\thanks{e-mail: rzhangab@connect.ust.hk}\\ %
     \scriptsize Hong Kong University of \\\scriptsize Science and Technology
\and Huamin Qu\thanks{e-mail: huamin@cse.ust.hk}\\ %
     \scriptsize Hong Kong University of \\\scriptsize Science and Technology}

\abstract{
With increasing popularity in online learning, a surge of E-learning platforms have emerged to facilitate education opportunities for k-12 (from kindergarten to 12th grade) students and with this, a wealth of information on their learning logs are getting recorded. However, it remains unclear how to make use of these detailed learning behavior data to improve the design of learning materials and gain deeper insight into students' thinking and learning styles. In this work, we propose a visual analytics system to analyze student learning behaviors on a K-12 mathematics E-learning platform. It supports both correlation analysis between different attributes and a detailed visualization of user mouse-movement logs. Our case studies on a real dataset show that our system can better guide the design of learning resources (e.g., math questions) and facilitate quick interpretation of students' problem-solving and learning styles.

}

\CCScatlist{
  \CCScatTwelve{Human-centered computing}{Visu\-al\-iza\-tion}{Visu\-al\-iza\-tion application domains}{Visual analytics}
}

\begin{document}


\maketitle

\section{Introduction}

With the increasing popularity of online education, many E-learning platforms have been launched to facilitate the education of K-12 students. The E-learning platforms are often equipped with various kinds of learning resources, such as item banks and videos. Rich learning logs such as grades, login time and mouse trajectories are often recorded When students interact with these learning materials.

Many success studies have been reported on using visual analytics to analyze students' learning behaviors based on learning logs. For example, Vismooc~\cite{shi2015vismooc} visualized the clickstream of video-watching behavior in MOOCs (Massive Open Online Courses); Viseq~\cite{chen2018viseq} analyzed the learning sequence of different learning activities (watching videos, doing exercise, etc.); PeerLens~\cite{xia2019peerlens} provided a visual interface to facilitate students planning their learning path through online question pools. However, few works have touched upon the analysis of detailed problem-solving steps within questions. It remains unclear how to make use of these rich learning behavior logs to improve the design of learning resources~\cite{costello2018evaluation} and better interpret students' thinking and learning styles.

In this work, we propose a visual analytics system to analyze students' learning behaviors on a K-12 mathematics E-learning platform from different levels of granularity. From a high level point of view, it supports a quick overview of student performance and learning behaviors on the whole question set. At the low level, it utilizes the heat map and the transition map to enable detailed exploration of problem-solving patterns on each question, and provide feedback and guidance to help educators optimize their teaching methods. We have also conducted three case studies, which shows that our system can better guide the design of learning materials (e.g., math questions) as well as facilitate quick interpretation of students' problem-solving and learning styles.

\section{Visual Analytics System}

\begin{figure}[h]
\centering
\includegraphics[width=0.85\linewidth]{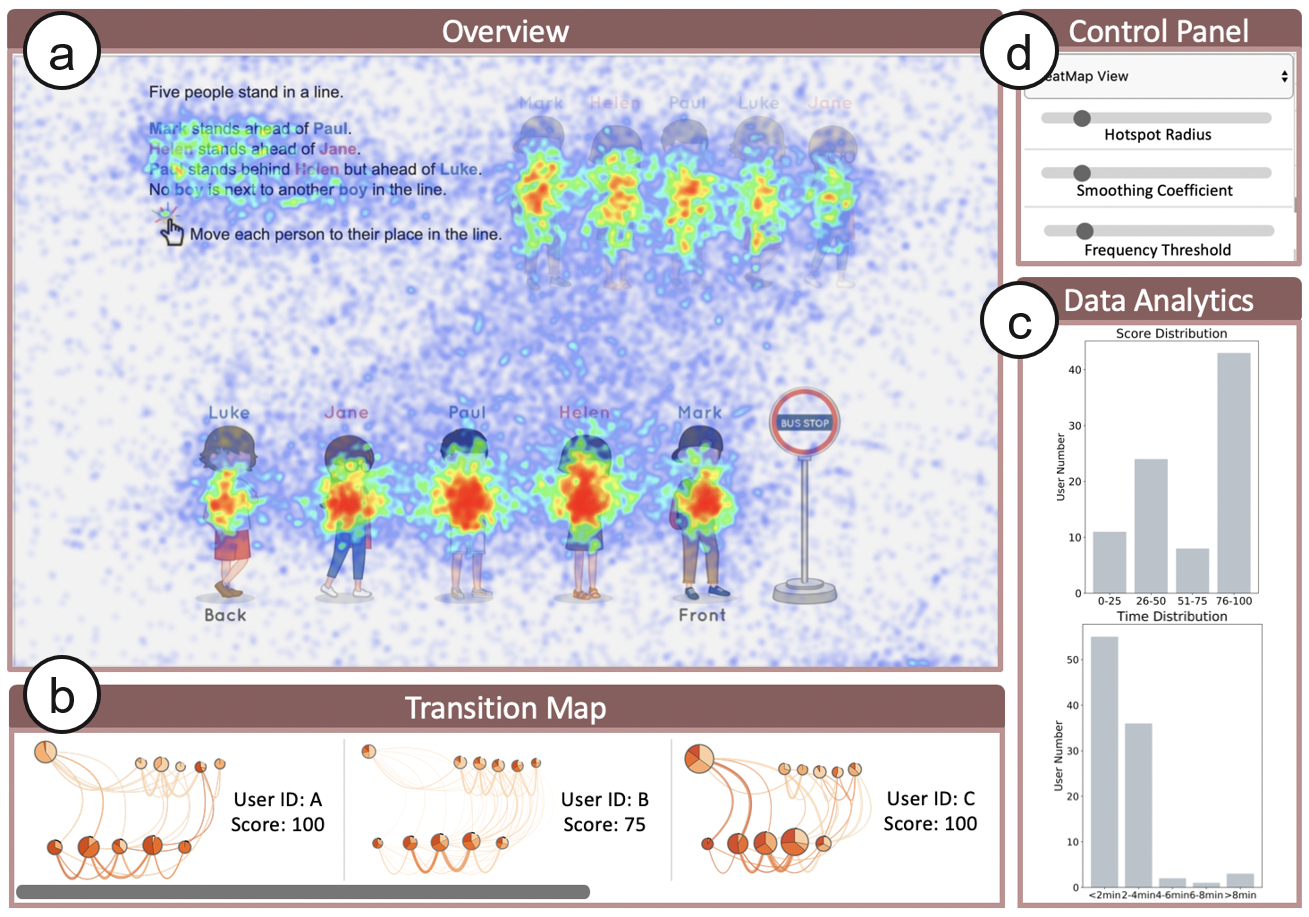}
\caption{The system interface.}
\centering
\label{fig:teaser}
 \vspace{-0.2cm}
\end{figure}

Fig.~\ref{fig:teaser} shows the system interface. It contains four views: (a) Overview of the problem and the heat map showing the problem-solving interaction data; (b) Transition map view revealing the detailed problem-solving steps; (c) Data analytics view displaying the correlation between learning attributes and learning performance; (d) Control panel for adjusting the size of Region of Interests (ROIs).

\begin{figure}[h]
\centering
\includegraphics[width=0.9\linewidth]{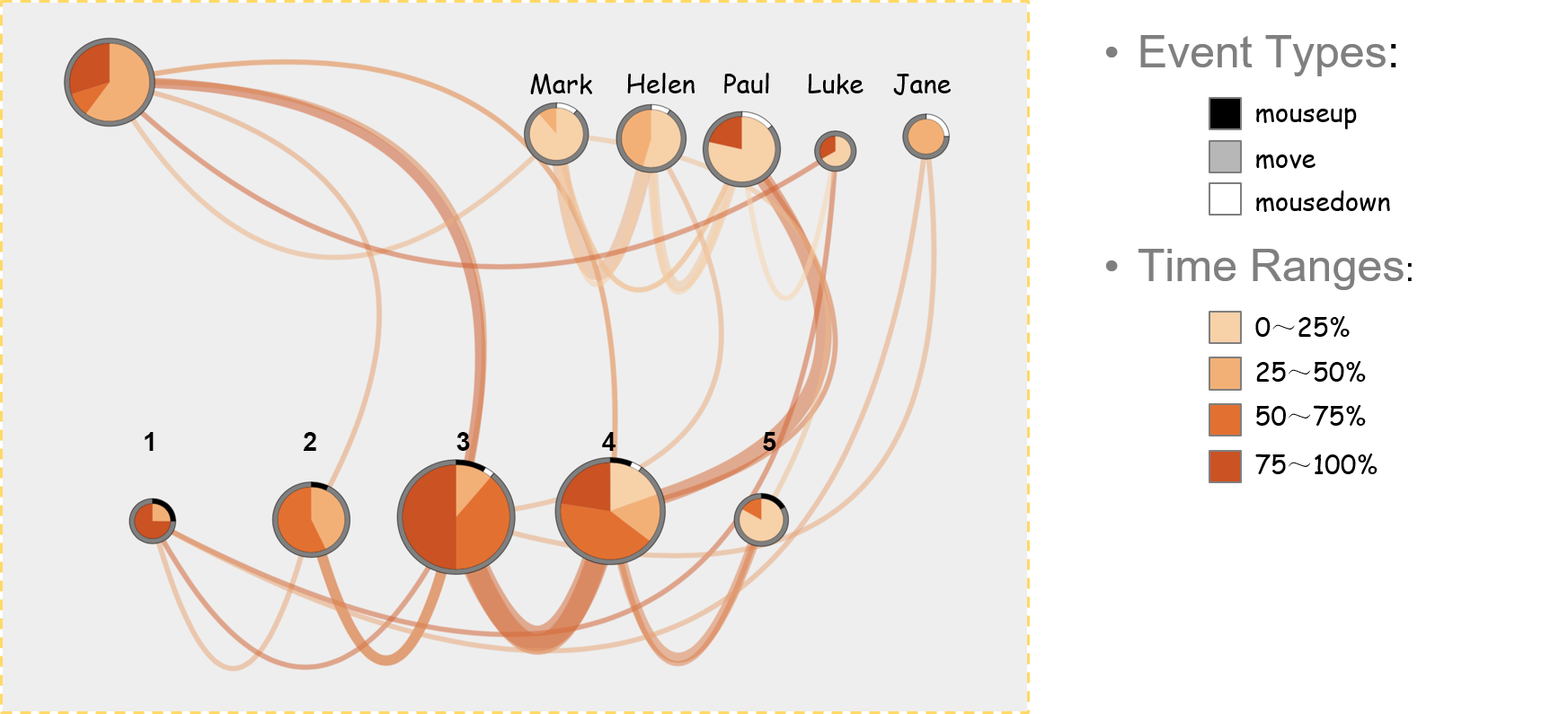}
\caption{The transition map demonstrating the detailed problem-solving steps with event types and time information.}
\centering
\label{fig:transitionmap}
\vspace{-0.5cm}
\end{figure}

As for the transition map in Fig~\ref{fig:transitionmap}, regions with intensive interaction are extracted as ROIs and visualized through pie charts, which are connected by arcs according to users’ mouse movement streams to form a transition map. The size of the pie charts represents the number of interactions over the ROIs; the color of the arcs and sectors represents the time order. In this way, the map reveals students’ problem-solving steps. We can see that the third and fourth pie at the bottom are larger than others and the lines between them are much thicker which means a large amount of transitions between these two ROIs.

\section{Case studies}

\subsection{Overall Performance Analysis}

\begin{figure}[h]
\centering
\includegraphics[width=0.5\linewidth]{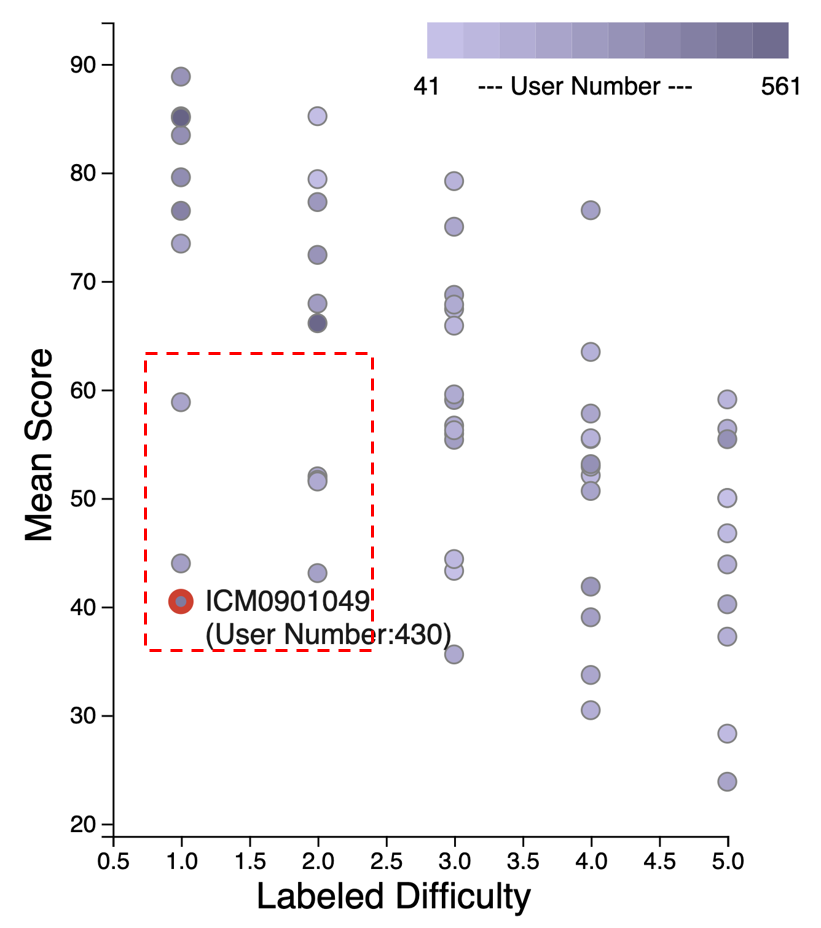}
\caption{Overview of student performances on all questions.}
\centering
\label{fig:c1}
\end{figure}

In this first case study, we use our system to perform a correlation analysis on the students' mean scores with the labeled difficulty of the questions.
Fig.~\ref{fig:c1} shows the correlation between the predefined difficulty of each question and the mean score of all students. It is worth mentioning that some questions labeled by the question designer as "easy" are actually having lower mean scores (highlighted in the dashed rectangle), which suggests that their difficulties should be re-evaluated.

\subsection{Thinking Patterns Identification}

\begin{figure}[h]
\centering
\includegraphics[width=1.0\linewidth]{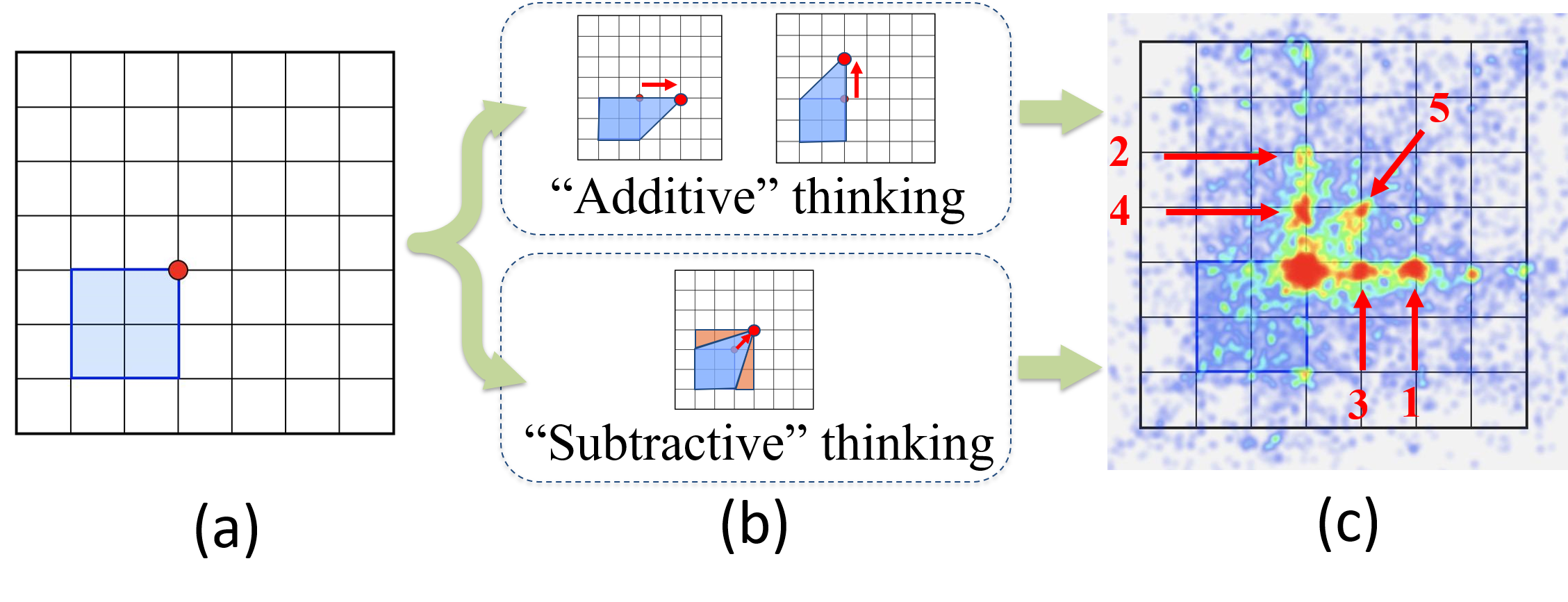}
\caption{(a) An example math problem: drag the red dot to make the area 6 boxes instead of 4. (b) Different thinking patterns in solving the problem. (c) A heat map showing the problem-solving interaction. }
\centering
\label{fig:c2}
\vspace{-0.5cm}
\end{figure}

The system can also reveal different thinking modes. As shown in Fig.~\ref{fig:c2}a, students are asked to drag the red dot to make a new shape with a area of 6 squares instead of 4. The two edges, left and bottom, are fixed and the red dot only changes the top and right edges. We analyzed the mouse movement logs through a heat map (Fig.~\ref{fig:c2}c) and identified the thinking patterns of the students. Point 1, 2 and 5 (indicated by red arrows on the Figure) are all possible answers. According to the heat map, many students solved the question in the “additive” way of thinking (Fig.~\ref{fig:c2}b) and preferred to move the point horizontally (passing through point 3 to point 1) rather than vertically (passing through point 4 to point 2). A small group of students also solved the question in a “subtractive” way and pinned at point 5.

\subsection{Problem-solving Steps Reveal}

\begin{figure}[h]
\centering
\includegraphics[width=0.8\linewidth]{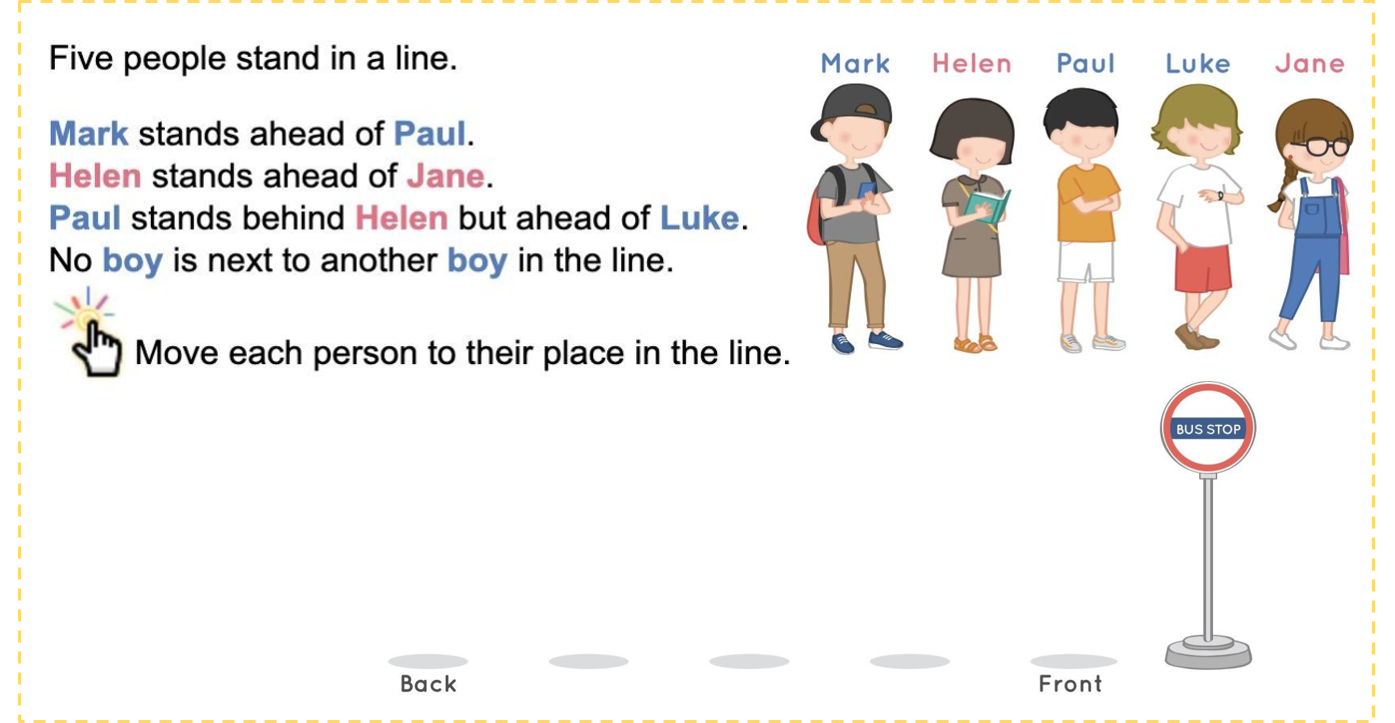}
\caption{Another example of the math problem.}
\centering
\label{fig:example}
 \vspace{-0.5cm}
\end{figure}

\begin{figure}[h]
\centering
\includegraphics[width=1.0\linewidth]{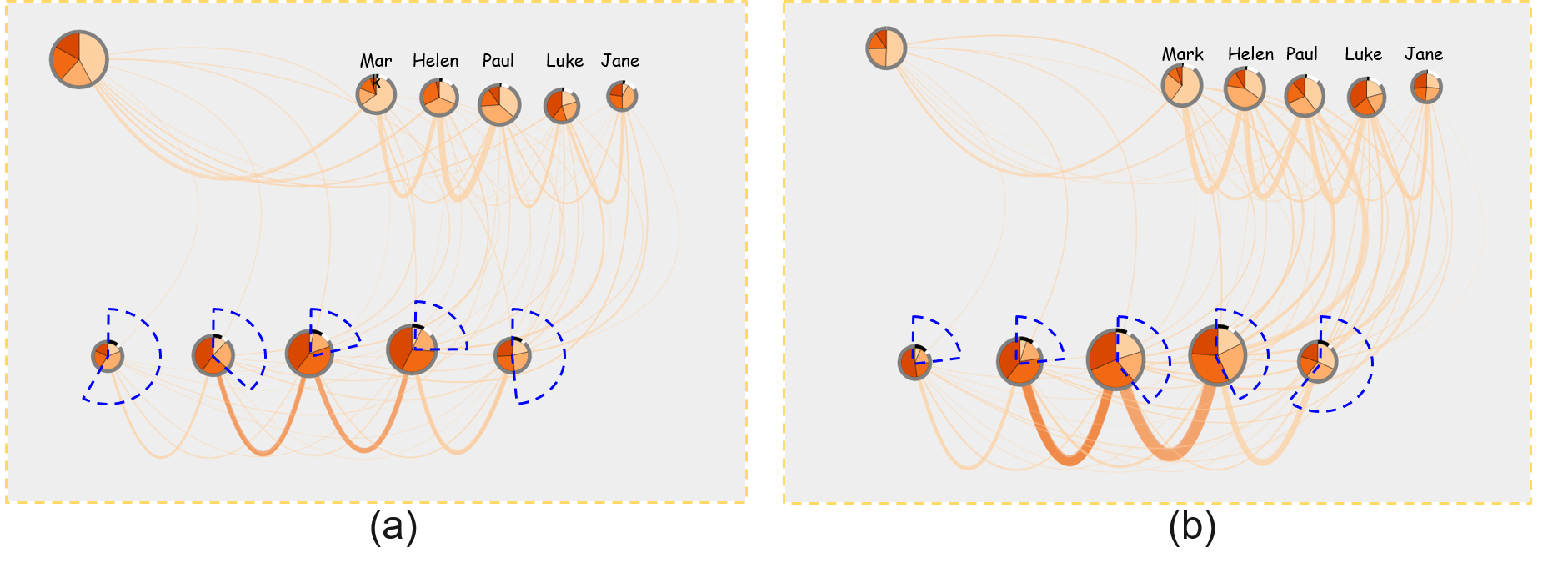}
\caption{(a) The transition map of students with wrong answers. (b) The transition map of students with full marks.}
\centering
\label{fig:c3}
 \vspace{-0.5cm}
\end{figure}

However, for questions like Fig.~\ref{fig:example}, there is only one unique answer. If solely using the heat map, the key steps to solve a problem cannot be inferred. Therefore, we use a transition map to reveal more details. In Fig.~\ref{fig:c3}, (a) is the transition map of students with wrong answers; (b) is the transition map of students with full marks. We can see that the light area of pies in (a) is decreasing from the left to the right while in (b) it is increasing. Since the color represents the time order of the sequence, a conclusion can be drawn that the students with full marks tend to order the people from right to left, while students with wrong answers seem to do it in the opposite way.

\section{Conclusion and future work}
In this poster, we proposed a visual analytics system to analyze student learning behaviors on K-12 math E-learning platforms. It enables both correlation analysis of different learning data attributes and visualization of detailed interactions of students. We evaluated the proposed system with real log data from a K-12 math E-learning platform. Our case studies show that the proposed system can help instructors quickly find the possible flaws in the design of learning materials (e.g., which question difficulty levels need to be revised), as well as gaining deeper insight into the detailed problem-solving styles of different students. In the future, we plan to further improve the current system by incorporating comprehensive analysis of fine-grained problem-solving behaviors (i.e., performance prediction).


\bibliographystyle{abbrv-doi}

\bibliography{template}
\end{document}